\documentclass[11pt]{article}
\usepackage{amssymb,amsmath,amsthm}
\theoremstyle{plain}

\theoremstyle{remark}

\def\bnumber#1{\textbf{\textsf{#1}}}

\numberwithin{thm}{section}

\begin{document}

%%\title

{Title: Can the Salecker--Wigner clock be microscopic?}

%%\author

{Author: Andor Frenkel (Institute for Particle and
Nuclear Physics, Budapest, Hungary)}

%%\date{}

%%\maketitle

Half a century ago H. Salecker and E. P. Wigner
examined the functioning of a quantum clock of very simple
construction [1].
They raised the question whether such a clock can be microscopic
or not, but no clear-cut answer has been reached in [1] (see also
[2]).
In this note it is shown that their clock can have a microscopic
mass and size only if its accuracy is poor, but if the size is
macroscopic, a decent accuracy can be achieved even if the mass
is microscopic.

\section*{Can the Salecker--Wigner clock be microscopic?}

\subsection*{1. Introduction}

In an article published in the Physical Review in 1958, H.
Salecker and E. P. Wigner emphasized that not only time-like,
but also space-like distances can be measured making use of
clocks only [1].
The possibility of avoiding the use of measuring rods is
important, because rods are essentially macroscopic objects,
whereas there is no principle which excludes the existence of
microscopic quantum clocks.
As pointed out in the first paragraph of [1], the use of such a
clock, when measuring distances between world events, allows to
assess the limitations imposed by quantum mechanics on these
basic measurements.

In Section 3 of [1] the authors have carried out a detailed
study of a quantum clock (called ``the S-W clock'' in this
note), of simple, realizable construction.
They have shown that the reading of this clock, in other words
the observation of the time kept by the clock, does not change
the order of magnitude of the limitations mentioned above. Also,
they established a formula determining the order of magnitude of
the minimal mass of the S-W clock, needed to achieve a
prescribed accuracy.

The title given in [1] to the section devoted to the S-W clock
reads ``Example of a simple microscopic clock.'' However, at the
end of the introductory Section 1 it is said that ``the clocks we
can construct are not wholly microscopic.'' Moreover, in article
[2], published by Wigner in 1957, the formula in question is
quoted, and a numerical example -- the only one in [1] and [2] --
is given, in which the mass of the clock turns out to be almost a
gram, the mass of an obviously macroscopic object. Accordingly,
referring to paper [1] then in preparation, Wigner writes that
``the result in summary is that a clock is an essentially
non-microscopic object.''

The purpose of the present note is to clarify this puzzling situation.
The conclusion will be that as far as its mass is concerned, the
simple clock designed in Section~3 of [1] can be macroscopic or
microscopic, depending on the choice of its parameters. However,
if the mass is microscopic, the linear size of the clock can be
restricted to a microscopic value, e.g.\ to a few Angstr\"oms,
only if the accuracy of the clock is poor.
This latter result is in agreement with the warning on page 261
of [2]: ``what we vaguely call an atomic clock, a single atom
which ticks off its periods, is surely an idealization which is
in conflict with fundamental concepts of measurability.''

\subsection*{2. Description of the S-W clock under the neglect
of quantum effects}

The simple quantum clock mentioned in the introduction consists of
three free bodies \bnumber{1, 2, 3} of equal masses $M$. Bodies
\bnumber 1 and \bnumber 3, at rest in an inertial frame of
reference at a distance $2\ell$ from each other, are the
ends of the linear dial which has no inner division points. Body
\bnumber{2}, moving freely with velocity $-u$ $(0 < u \ll c)$ from
the end \bnumber{3} of the dial to the other end, is the hand.
(See Fig. 1. As a rule, the conventions and the notations used in
[1] have been retained. However, the notation $v$ for the negative
velocity of the hand has been replaced by $-u$. Notice that in Eq.
(9) of [1] the denominator should be not $v$, but $|v|$.)
The time
interval $T$ during which the hand arrives from \bnumber{3} to
\bnumber{1} is the running time of the clock. Obviously
\begin{equation}
T = \frac{2\ell}{u};
\label{eq.1}
\end{equation}
the running time cannot be made several times $T$ by letting the
hand scatter off and oscillate between the dial bodies, because
the latter ones are not fixed.

\begin{figure}
\begin{alignat*}{4}
&1 & \hskip30.5mm & & \hskip4.5mm
&\raisebox{-5pt}{$-u$}\hskip7.5mm 2  &
\hskip10mm &3 \\[-4.5mm]
&\text{\huge$\bullet$} & &\raisebox{5pt}
{\tiny$|$} &
&\phantom{-u}\text{\huge$\boldsymbol\leftarrow\!\!\!\bullet$} &
&\text{\huge$\bullet$}\phantom{\hskip10mm} \\[-6.55mm]
\noalign{\hspace*{13mm}\hbox to101mm{\rightarrowfill}}\\[-15mm]
& & & & & & &\hskip 15.5mm X\\[-6.5mm]
-&\ell & &\hspace*{2pt} 0 & & & &\ell \\[-4mm]
&t_h = +\frac{T}{2} & &\hspace*{2pt} t_h = 0 & & & &t_h = -\frac{T}{2}
\end{alignat*}

\centerline{Fig. 1}
\end{figure}

\subsection*{3. The relation between the mass and the accuracy of
the S-W clock}

The time $t_h$ kept by a clock having a hand depends on the
position $x_h$ of the hand relative to the dial. In the case of a
quantum clock the hand is a quantum body, and its center of mass
coordinate $x_h$ has an indeterminacy $\Delta x_h$. For a free
body there is, in principle, no obstacle to take for its center of
mass wave function a (nearly) minimal Gaussian wave packet. Then
$\Delta x_h$ is the width of the packet, and
\begin{equation}
\Delta x_h \cdot \Delta p_h \approx \hslash ,
\label{eq.2}
\end{equation}
where $-p_h$ is the momentum of the hand.
It should be noted here that in [1] the S-W clock is studied in
the framework of a spacetime which, in addition to the time-like
direction, has only one space-like dimension.
This restriction has been maintained in the present note,
although, as noticed in the conclusion, it can be lifted.
The symbol ``$\approx$'' in \eqref{eq.2} and below stands for
``equal, at least in order of magnitude.''

The accuracy of a clock is characterized by the time interval
$\tau$ which the clock already cannot resolve.
The higher the accuracy, the smaller $\tau$ should be.
Also, for a good clock $\tau$ should be much smaller than the
running time $T$, in other words the relative accuracy
\begin{equation}
n = \frac{T}{\tau}
\label{eq.3}
\end{equation}
should be much larger than $1$,
\begin{equation}
n \gg 1.
\label{eq.4}
\end{equation}

The accuracy of a quantum clock is in fact the uncertainty
$\Delta t_h$ in the moment of time of the passage of the center
of mass wave packet of the hand through a fixed point of the dial.
Since the hand moves with velocity of magnitude $u$,
\begin{equation}
\Delta t_h = \frac{\Delta x_h}{u}.
\label{eq.5}
\end{equation}
$\Delta x_h$, and therefore $\Delta t_h$, too, will keep their
order of magnitude during the whole running time $T$, if the
initial value of $\Delta x_h$ satisfies the relation
\begin{equation}
T \approx \frac{M(\Delta x_h)^2}{\hslash }.
\label{eq.6}
\end{equation}
Indeed, as is well known, this relation guarantees that a
free Gaussian wave packet will only double its width $\Delta
x_h$ during the running time $T$.
Thus, if $\Delta t_h = \tau$ holds at the beginning of the run,
\begin{equation}
\Delta t_h \approx \tau
\label{eq.7}
\end{equation}
will hold during the whole run, the order of magnitude of the
accuracy of the S-W clock remains $\tau = T / n$.

From \eqref{eq.1}, \eqref{eq.3} and \eqref{eq.5} it follows that
\begin{equation}
n = \frac{T}{\tau} \approx \frac{2\ell}{\Delta x_h},
\label{eq.8}
\end{equation}
and from \eqref{eq.2}, \eqref{eq.6}, \eqref{eq.8} and
\eqref{eq.1} one sees that
\begin{equation}
\Delta u = \frac{\Delta p_h}{M} \approx
\frac{\hslash }{M \Delta x_h} \approx \frac{\Delta x_h}{T} \approx
\frac{2\ell}{T} \cdot \frac1{n} = \frac{u}{n} \ll u,
\label{eq.9}
\end{equation}
as it should be.

Expressing $\Delta x_h$ in \eqref{eq.6} with the help of
\eqref{eq.8}, one finds the dependence of $M$ on $2\ell$, $T$ and
$n$ (or on $\tau$):
\begin{equation}
M \approx \frac{\hslash  T}{(2\ell)^2} n^2 = \frac{\hslash  T^3}{(2\ell)^2
\tau^2} .
\label{eq.10}
\end{equation}
The mass of the clock is not $M$ but $3M$, but as far as orders
of magnitudes are concerned, one can forget about this difference.

The dial bodies \bnumber{1} and \bnumber{3} are quantum bodies, too.
It is easy to see that if the indeterminacies in their positions
are taken equal to that of the hand, then relation \eqref{eq.10}
remains valid.

\subsection*{4. The minimal mass of an S-W clock needed to
achieve a desired accuracy}

According to \eqref{eq.10}, for a given running time $T$ and for
an accuracy $\tau$ to be achieved, the mass of the S-W clock is
inversely proportional to $(2\ell)^2$.
The minimal value of the mass of the S-W clock is therefore
determined by the maximal value of the length of the dial.

Relying on simplicity, and on a plausibility argument exposed on
page 261 of [2] before and after Eq. \eqref{eq.2}, Salecker and
Wigner take the maximal length of the dial equal to the distance
covered by light during the time interval $\tau$:
\begin{equation}
2\ell_{\max} \approx c\tau.
\label{eq.11}
\end{equation}
From \eqref{eq.10} one finds then that
\begin{equation}
M_{\min} \approx \frac{\hslash }{c^2 \tau} n^3; \quad \left(n =
\frac{T}{\tau} \gg 1\right).
\label{eq.12}
\end{equation}
In what follows, the length of the dial and the mass of the
clock will always be taken at their extremal values, and the
indices ``max'' and ``min'' will be omitted.
Comparing the expression
\begin{equation}
2\ell \approx c\tau
\label{eq.13}
\end{equation}
for the maximal value of the dial with \eqref{eq.1}, one finds that
\begin{equation}
u \approx \frac{c}{n} \ll c,
\label{eq.14}
\end{equation}
as it should be for the velocity $u$ in non-relativistic quantum
mechanics.

\subsection*{5. Requirements to be met by an S-W clock}

Before looking at the requirements, let us list the mathematical
relations to be used.
One of them is the formula for the minimal mass,
\begin{equation}
M \approx \frac{\hslash }{c^2 \tau} n^3 \approx 10^{-48} \frac{n^3}{\tau},
\label{eq.15}
\end{equation}
where in the last term the value of $\hslash  / c^2$ is given in CGS
(Centimeter, Gram, Second) units.
If not specified otherwise, these units will be used in what
follows.

\eqref{eq.8} and \eqref{eq.14} imply that
\begin{equation}
n = \frac{T}{\tau} \approx \frac{2\ell}{\Delta x} \approx
\frac{c}{u},
\label{eq.16}
\end{equation}
where $\Delta x$ stands for the equal widths of the wave packets
of the hand and of the dial bodies, and, of course,
\begin{equation}
n \gg 1.
\label{eq.17}
\end{equation}
Owing to \eqref{eq.5} and \eqref{eq.7}
\begin{equation}
\Delta x \approx u\tau.
\label{eq.18}
\end{equation}
For given values of $\tau$ and $n$ (or of $\tau$ and $T$), all
the quantities in \eqref{eq.15}, \eqref{eq.16}, \eqref{eq.17}
and \eqref{eq.18} are determined.
Two more quantities to be considered are the Compton wavelength
$L_M$ corresponding to the mass $M$,
\begin{equation}
L_M = \frac{\hslash }{Mc} \approx \frac{10^{-38}}{M},
\label{eq.19}
\end{equation}
and the radius $R$, equal for the hand and of the dial bodies,
\begin{equation}
R = 0.62 \left(\frac{M}{\varrho}\right)^{1/3},
\label{eq.20}
\end{equation}
where $\varrho$ is the density of the bodies \bnumber{1, 2, 3}
in three-dimensional space.
A comment on the departure from the one-dimensional space model
when considering $\varrho$ is given in the last section of this note.
In [1] the  size and the density of the hand and of the dial
bodies were not discussed.

The requirements to be satisfied by a non-relativistic,
microscopic S-W clock are

a) The relation
\begin{equation}
n = \frac{T}{\tau} \gg 1
\label{eq.21}
\end{equation}
should hold for any decent clock.

b)
\begin{equation}
L_M \ll \Delta x
\label{eq.22}
\end{equation}
should hold for any quantum object in non-relativistic quantum
mechanics.
Inserting the expression for $M$ given in \eqref{eq.15} into
\eqref{eq.19} and relying on \eqref{eq.14} and \eqref{eq.18} one
finds that
\begin{equation}
L_M \approx \frac{\Delta x}{n^2},
\label{eq.23}
\end{equation}
so that requirement b) follows from a).

c) It is a natural requirement that the geometrical size $R$ of
the bodies constituting the S-W clock be much smaller than the
length of the dial:
\begin{equation}
R \ll 2\ell .
\label{eq.24}
\end{equation}

d) An important characteristic of the microbehavior of a
quantum body is that the indeterminacy in the position of its
center of mass is much larger than its geometrical size:
\begin{equation}
\Delta x \gg R.
\label{eq.25}
\end{equation}
Macrobehavior is characterized by the opposite property: the
indeterminacy in the localization of the center of mass is much
smaller than the geometrical size of the body.

Since the fulfillment of requirement b) follows from a) which
will be supposed to hold always, only requirements c) and d)
should be verified for the numerical examples considered below.

\subsection*{6. Example of a macroscopic S-W clock}

On page 261 of [2] Wigner has considered a quantum clock with
accuracy
\begin{equation}
\tau = 10^{-8} \text{ sec},
\label{eq.26}
\end{equation}
and with running time
\begin{equation}
T = 1 \text{ day } = 8.64 \cdot 10^4 \text{ sec},
\label{eq.27}
\end{equation}
i.e. a clock with relative accuracy
\begin{equation}
n = 8.64 \cdot 10^{12}.
\label{eq.28}
\end{equation}
The mass of this clock is
\begin{equation}
M \approx 0.072 \text{ gram},
\label{eq.29}
\end{equation}
an obviously macroscopic value.
The length of the dial is macroscopic, too,
\begin{equation}
2 \ell \approx c\tau = 3 \text{ meter},
\label{eq.30}
\end{equation}
while the width of the wave packet of the hand (and of the dial
bodies) is small even on the atomic scale:
\begin{equation}
\Delta x \approx \frac{2\ell}{n} \approx 10^{-11} \text{ cm}.
\label{eq.31}
\end{equation}
For usual terrestrial densities $\varrho \approx  1\text{
gram/cm}^3$, the radius $R$ of the hand is
\begin{equation}
R \approx 0.26 \text{ cm}.
\label{eq.32}
\end{equation}
Thus, the relations
\begin{equation}
R \ll 2\ell,
\label{eq.33}
\end{equation}
\begin{equation}
\Delta x \ll R
\label{eq.34}
\end{equation}
hold, the second one corroborating the macroscopic character of
the clock.
Notice that this clock is macroscopic both by its mass and its size.

\subsection*{5. Microscopic S-W clocks}

\subsubsection*{5.1. An S-W clock of microscopic mass, but of
macroscopic size}

The factor on the right-hand side of \eqref{eq.15}, leading to a
macroscopic value of $M$, is $n^3$, where the relative accuracy
$n$ itself should be much larger than 1.
In the example given in [2] and recalled above, $n$ is almost
$10^{13}$.
A clock with a much smaller relative accuracy, e.g.\ with
\begin{equation}
n = 10^7,
\label{eq.35}
\end{equation}
is still a decent clock.
Let us look at such a clock, with
\begin{equation}
\tau = 10^{-7} \text{ sec},
\label{eq.36}
\end{equation}
\begin{equation}
T = 1 \text{ sec}.
\label{eq.37}
\end{equation}
Then \eqref{eq.15} gives
\begin{equation}
M \approx 10^{-20} \text{ gram } \approx 10^4 \cdot M_N,
\label{eq.38}
\end{equation}
where $M_N$ is the mass of the nucleon.
The hand and the dial bodies consist of $10^4$ nuclei, or $10^4
\div 10^3$ light atoms.
10~000 nuclei do not form a stable nucleus, but from the atoms
solid bodies can be constructed.
At normal terrestrial densities $\varrho \approx 1\text{
gram/cm}^3$, $10^{-20}$ gram corresponds to a radius
\begin{equation}
R \approx 10^{-7} \text{ cm}.
\label{eq.39}
\end{equation}
The velocity of the hand is
\begin{equation}
u \approx \frac{c}{n} = 3\cdot 10^3\ \frac{\text{cm}}{\text{sec}}\, ,
\label{eq.40}
\end{equation}
so that
\begin{equation}
\Delta x \approx u \tau \approx 3 \cdot 10^{-4} \text{ cm},
\label{eq.41}
\end{equation}
and the requirement
\begin{equation}
\Delta x \gg R
\label{eq.42}
\end{equation}
for microscopic behavior is fulfilled, corroborating the intuitive
feeling that the quantum properties of a body containing only
10~000 atoms cannot be ignored. However, similarly to the  clock
in [2], the length of the dial is macroscopic:
\begin{equation}
2\ell \approx n \cdot \Delta x \approx 30 \text{ meter}.
\label{eq.43}
\end{equation}
In the next subsection the possibility of designing S-W clocks
with microscopic mass and size is examined.

\subsubsection*{5.2. Difficulties in designing S-W clocks of
microscropic mass and size}

A survey of the possibilities of designing an S-W clock of
microscopic mass $M \leq 10^{-16} \text{ gram}$ and microscopic
size $2\ell \leq 10^{-5} \text{ cm}$ reveals that only clocks
with low relative accuracy $n \leq 100$ satisfy the requirements
\begin{equation}
R \ll 2\ell
\label{eq.44}
\end{equation}
and
\begin{equation}
R \ll \Delta x.
\label{eq.45}
\end{equation}
Instead of a tedious general survey, a few examples will be
presented below.

Let us look first at the case when
\begin{equation}
n = 100,
\label{eq.46}
\end{equation}
and
\begin{equation}
M \approx M_N = 10^{-24} \text{ gram}.
\label{eq.47}
\end{equation}
In CGS units one finds
\begin{equation}
\tau \approx 10^{-18}, \quad T \approx 10^{-16},
\label{eq.48}
\end{equation}
\begin{equation}
\Delta x \approx 3 \cdot 10^{-10}, \quad 2\ell \approx 3 \cdot 10^{-8},
\label{eq.49}
\end{equation}
\begin{equation}
u \approx 3 \cdot 10^8 .
\label{eq.50}
\end{equation}
Now one cannot take a light atom (e.g. a hydrogen or a helium
atom) instead of a nucleon or of a light nucleus, because for an
atom $R \approx 10^{-8}\text{ cm}$, and the requirements
\begin{equation}
R \ll 2\ell, \quad R \ll \Delta x
\label{eq.51}
\end{equation}
would not be met.
The radius of a nucleon, or of a light nucleus,
\begin{equation}
R \approx 10^{-13} \text{ cm}
\label{eq.52}
\end{equation}
is an acceptable value.
The short, $10^{-13}\text{ cm}$ range strong interaction between
such bodies along the dial of $3\cdot 10^{-8}\text{ cm}$ can be
safely neglected, but the effect of the Coulomb repulsion
between the nuclei, due to their protons, should be estimated;
the bodies \bnumber{1, 2, 3} of the clock are not completely free.
The simplest choice would be to construct the clock out of three
protons.
However, the procedure for reading the S-W clock, proposed in
[1], requires that the bodies constituting the clock be
distinguishable from each other through their interaction with
photons.
Therefore, instead of protons, one should take three light
nuclei of (nearly) equal masses in the range of $10^{-24} \div
10^{-23}$ gram, but with different $\gamma$-ray spectra.

So, S-W clocks of microscopic mass and size can be designed, but
their relative accuracy $T/\tau$ is only 100. Notice that the
velocity of the hand of such clocks is $c/100$, which is not a
comfortably non-relativistic velocity.

The choice of the nucleon mass for $M$ is not the only possible
choice for $n = 100$, but the range of the acceptable mass
values is not very wide.
For $M \geq 10^{-20}$ gram and $M \leq 10^{-27}$ gram either at
least one of the conditions \eqref{eq.24}, \eqref{eq.25} is
violated, or the length of the dial is macroscopic, $2\ell \geq
10^{-5} \text{ cm}$.

For $n = 10^3$ and $n = 10^4$ the mass $M \approx 10^{-20}
\text{ gram} \approx 10^4 M_N$ would be acceptable, if a nucleus
containing 10~000 nuclei would be stable, but this is not the case.
For $n \geq 10^5$, and for any non-macroscopic mass $M \leq
10^{-16}$ gram, again either at least one of the two
requirements \eqref{eq.24}, \eqref{eq.25} is violated, or the
length of the dial is macroscopic.

If $n = 10$, the requirements \eqref{eq.24}, \eqref{eq.25} are
met for $M \approx M_N$, but the situation is worse than in the
already not too good case with $n = 100$.
Indeed, now $u = 0.1 c$, and while for $n = 100$ one found
$\Delta x \approx 10^3 R$, here $\Delta x \approx 10 R$ only.

\subsection*{6. Conclusion and outlook}

It has been shown in the preceding two sections that microscopic
S-W clocks can be designed if the relative accuracy $n = T/c$ is
not excessively large.
For $n = 10^7$ an S-W clock of microscopic mass $M \approx
10^{-20}$ gram, but of macroscopic dial length $2\ell \approx
30$ meter has been described.
The macroscopic separation between microscopic objects is not a
reason for ignoring the quantum properties of those objects.
It has been also shown that for low relative accuracy $n \leq
100$ the length of the dial can be made microscopic, too.

It should be noted that if one accepts the very large relative
accuracy $n \approx 10^{13}$ advocated in the example given in
[2], and if one wishes to design an S-W clock of microscopic
mass (with $M \leq 10^{-16}\text{ gram}$), one finds that the
running time would be by many orders of magnitude larger than
the age of the Universe.
Microscopic S-W clocks cannot have such a high relative accuracy.

It will be shown in a forthcoming paper that for the correct
functioning of the reading mechanism proposed in [1], the
maximal length of the dial should be indeed $c\tau$, a claim
made in [1] on the basis of a plausibility argument.
It will also be shown that the results obtained in [1] in the
framework of the $1 + 1$ dimensional spacetime remain valid in
the physical $1 + 3$ dimensional spacetime.
This is why in the preceding section realistic volume densities
have been considered instead of linear ones.

As noticed on page 574 of [1], it is possible that quantum
clocks ``more clever'' than the S-W clock can be found.
If such clocks do exist, then for given accuracy and running
time the value of their minimal mass should be smaller than that
of the S-W clock.

\subsection*{7. Acknowledgement}

I am indebted to Istv\'an R\'acz, Abner Shimony and L\'aszl\'o
Szabados for their interest in this work, and for the careful
reading of the manuscript.
This research was partially supported by OTKA grant T~034337.


\begin{thebibliography}{2}

\bibitem[1]{} H. Salecker and E. P. Wigner,
Phys. Rev. \emph{109}, 571, 1958.


\bibitem[2]{} E. P. Wigner,
Rev. Mod. Phys. \emph{29}, 255, 1957.

\end{thebibliography}
\end{document}